\documentstyle[preprint,eqsecnum,aps]{revtex}

\begin{document}

\draft

\title{Canonical and D-transformations in Theories
with Constraints}

\author{Dmitri M. Gitman}

\address{Instituto de F\'{\i}sica, Universidade de S\~ao Paulo \\
P.O. Box 66318, 05389-970-S\~ao Paulo, S.P., Brasil}

\date{\today}

\maketitle

\begin{abstract}
We describe a
class of transformations in a super phase space (we call them
D-transformations), which play in theories with
second-class constraints the role of ordinary canonical
transformations in theories without constraints. Namely, in
such theories they preserve the
forminvariance of equations of motion, their quantum analogue
are unitary transformations, and the measure of integration in
the corresponding hamiltonian path integral is invariant under
these transformations.
\end{abstract}

\section{Introduction}
As is well known canonical transformations play an important role in
the hamiltonian
formulation of classical mechanics without constraints \cite{LaL}.
They preserve the forminvariance of the hamiltonian equations of motion
and their quantum analogue are unitary transformations
\cite{Dir1,Wey}. Canonical transformations constitute also a powerful
tool of the classical mechanics, which   allows one often to simplify
solution of the theory. For example, it is enough to mention that
evolution is also a canonical transformation. Quantum implementation
of canonical transformations where discussed in numerous papers, see
for example \cite{Wit,Itz,MoQ,And}. However, modern physical
theories in their classical versions are mostly singular (in
particular, gauge) ones, which means that in the hamiltonian formulation
they are theories with constraints \cite{Dir2,GiT1}. Equations of a
hamiltonian theory
with constraints are not form invariant under canonical
transformations, but namely this circumstance allows one to use these
transformations
to simplify the equations and to clarify the structure of the gauge
theory in hamiltonian formulation. Moreover,
formulations of a gauge theory in two diffrent gauges are connected by
means of a
canonical transformation \cite{GiT1,GiT2}. In general case, equations
of constraints change their form  ander the canonical transformations.
That is an indirect indication that the quantum version of the
canonical transformations in constrained theories is not an unitari
transformation (Of course, we are speaking about the complete
theory, but not about
its reduced unconstrained version). Thus, in case of constrained
theories one can believe that besides of the canonical transformation
another kind of transformations has to exist, which preserves the form
invarians of the equations of motion and which induces unitary
transformations on the quantum level. Namely they play the role of
ordinary  canonical transformations in theories without constraints.

In this paper we describe such kind of transformations for theories with
second-class constraints, which is, in fact, a general case, because of a
theory with first-class constraints can be reduced to a theory with
second-class ones by a gauge fixing. We call such transformation
D-transformations.

\section{Generalized canonical transformations}
Let a classical mechanics be given with phase variables $\eta
=(\eta^A),\;\;A=1,\ldots,2n$ (in general case they belong to Berezin
algebra \cite{Ber,GiT1} and have the Grassmann parities
$P(\eta^A)=P_A$), and with a symplectic metrics
$\Lambda^{AB}(\eta)$, which defines a generalized super Poisson bracket
for any   two
functions $F(\eta)$ and $G(\eta)$ with definite Grassmann parities
$P(F)$ and $P(G)$,
\begin{equation}\label{a1}
\left\{F,G\right\}^{(\eta,\Lambda)}=\frac{\partial_r F}{\partial \eta^A}
\Lambda^{AB}(\eta)\frac{\partial_l G}{\partial \eta^B }\;,
\end{equation}
where
$\partial_r/\partial\eta^A$ and
$\partial_l/\partial\eta^B$
are the  right and left derivatives respectively. The  metrics
$\Lambda^{AB}(\eta)$ is a $T_2$-antisymmetric  supermatrix \cite{GiT1},
$P(\Lambda^{AB})=P_A+P_B,\;\; \Lambda^{AB}(\eta)=-(-1)^{P_A
P_B}\Lambda^{BA}(\eta)$,
obeying the conditions,
\begin{equation}\label{a2}
(-1)^{P(A)P(C)}\Lambda^{AD}(\eta)\frac{\partial_l
\Lambda^{BC}(\eta)}{\partial \eta^D}+ {\rm cycl.}(A,B,C)=0\;,
\end{equation}
which are necessary and sufficient for the bracket (\ref{a1})
to be super antisymmetric and satisfy the super Jacobi identity,
\begin{eqnarray}\label{a3}
&&\left\{F,G\right\}^{(\eta,\Lambda)}=-(-1)^{P(F)P(G)}\left\{G,F\right\}^
{(\eta,\Lambda)}\;,\nonumber \\
&&(-1)^{P(F)P(K)}\left\{\left\{F,G\right\}^{(\eta,\Lambda)},K\right\}^
{(\eta,\Lambda)}+{\rm cycl.}(F,G,K)=0\;.
\end{eqnarray}
Besides, the property takes place
\begin{equation}\label{a4}
\left\{F,GK\right\}^{(\eta,\Lambda)}=\left\{F,G\right\}^{(\eta,\Lambda)}K+
(-1)^{P(F)P(G)}G\left\{F,K\right\}^{(\eta,\Lambda)}\;.
\end{equation}
It is easily to see that
\begin{equation}\label{a5}
\Lambda^{AB}(\eta)=\left\{\eta^A,\eta^B \right\}^{(\eta,\Lambda)}\;.
\end{equation}
In case if
\[
\Lambda^{AB}=E^{AB}=\left(
\begin{array}{rr}
0 & I \\
-I & 0
\end{array}
\right)\;,
\]
the generalized Poisson bracket (\ref{a1}) coincides with the ordinary
super Poisson bracket,
\begin{equation}\label{a6}
\left\{F,G\right\}^{(\eta,E)}=\frac{\partial_r F}{\partial \eta^A}
E^{AB}\frac{\partial_l G}{\partial \eta^B }=\left\{F,G\right\}\;.
\end{equation}

If $\eta '=\eta '(\eta)$ is a nonsingular change of variables, then
the generalized Poisson bracket (\ref{a1}) acquires in
the primed  variables the following form
\begin{equation}\label{a7}
\left\{F,G\right\}^{(\eta,\Lambda)}=\left\{F',G'\right\}^{(\eta ',\Lambda
')}=\frac{\partial_r F'}{\partial \eta'^A}
\Lambda'^{AB}(\eta')\frac{\partial_l G'}{\partial \eta'^B }\;,
\end{equation}
where
\begin{eqnarray}\label{a8}
&&F'(\eta ')=F(\eta )\;,\;\;\; G'(\eta ')=G(\eta)\;,\nonumber \\
&&\Lambda'^{AB}(\eta ')=\frac{\partial_r \eta'^A}{\partial \eta^C}
\Lambda^{CD}(\eta)\frac{\partial_l \eta'^B}{\partial \eta^D}
=\left\{\eta '^A,\eta '^B \right\}^{(\eta,\Lambda)}\;.
\end{eqnarray}

By analogy with the case of the ordinary Poisson bracket  one can ask
the question:
which kind of transformations preserves the  generalized Poisson
bracket  forminvariant, namely when a relation holds
\begin{equation}\label{a9}
\Lambda'^{AB}(\eta')=\Lambda^{AB}(\eta')\;.
\end{equation}
We will call
such kind of transformations  generalized canonical ones. They
are just canonical transformations in case when the generalized
Poisson bracket coincides with the ordinary Poisson bracket.

Consider transformations of the form
\begin{equation}\label{a10}
\eta '=e^{\check{W}}\eta\;.
\end{equation}
In (\ref{a10}) the operator $\check{W}$ is defined by its action on
functions of $\eta$,
\begin{equation}\label{a11}
\check{W}F(\eta)=\left\{F ,W \right\}^{(\eta,\Lambda)}\;,
\end{equation}
where $W(\eta),\;\; (P(W)=0)$, is a generating function of the
transformation. We are
going to demonstrate that the transformations (\ref{a10}) are just
the generalized canonical transformations, connected continuously with
the identical transformation. To this end one has, first,
to verify that the following property takes place
\begin{equation}\label{a12}
e^{\check{W}}F(\eta)=F(e^{\check{W}}\eta)=F(\eta')\;.
\end{equation}
Indeed, one can see, using (\ref{a4}), that
\begin{equation}\label{a13}
e^{\check{W}}F(\eta)e^{-\check{W}}=\sum_{n=0}^{\infty}\frac{1}{n!}[\check{W},[
\check{W},\ldots,
[\check{W},F]\ldots]]=e^{\check{W}}F(\eta)\;.
\end{equation}
Then, one can write, for example, for any analytic function $F(\eta)$,
\[
e^{\check{W}}F(\eta)=e^{\check{W}}F(\eta)e^{-\check{W}}=F\left(e^{\check{W}}\eta
e^{-\check{W}}\right)=F\left(e^{\check{W}}\eta\right)=F(\eta')\;.
\]
Now, let us introduce a function $F^{AB}(\alpha,\eta),\;\;P(\alpha)=0$,
\begin{equation}\label{a14}
F^{AB}(\alpha,\eta)=\left\{e^{\alpha\check{W}}\eta^A,e^{\alpha\check{W}}\eta^B
\right\}^{(\eta,\Lambda)}\;.
\end{equation}
At $\alpha=0$ this function coincides with $\Lambda^{AB}(\eta)$, see
(\ref{a5}), and at
$\alpha=1$ with $\Lambda'^{AB}(\eta')$, see (\ref{a8}) and (\ref{a10}),
\begin{eqnarray}
&&F^{AB}(0,\eta)=\Lambda^{AB}(\eta)\;,  \label{a15}\\
&&F^{AB}(1,\eta)=\Lambda'^{AB}(\eta')\;.\label{a16}
\end{eqnarray}
Differentiating (\ref{a14}) with respect to $\alpha$ and using the
Jacoby identity (\ref{a3}), one can get an equation for the function
$F^{AB}(\alpha,\eta)$,
\begin{equation}\label{a17}
\frac{\partial F^{AB}(\alpha,\eta)}{\partial \alpha}=\check{W}F^{AB}
(\alpha,\eta)\;.
\end{equation}
A solution of this equation, which obeys the initial condition
(\ref{a15}), has the form
\begin{equation}\label{a18}
F^{AB}(\alpha,\eta)=e^{\alpha\check{W}}\Lambda^{AB}(\eta)\;.
\end{equation}
Taking into account the equation
(\ref{a16}) and the property (\ref{a12}), we get just the condition
(\ref{a9}) of the forminvariance of the generalized Poisson bracket.
Thus, the transformations (\ref{a10}) are namely
generalized canonical transformations, connected continuously with the
identical transformation. By  definition they preserve the
forminvariance of the generalized Poisson bracket,
\begin{equation}\label{a19}
\left\{F,G\right\}^{(\eta,\Lambda)}=\left\{F',G'\right\}^{(\eta ',\Lambda
)}\;,\;\;\;F'(\eta ')=F(\eta ),\; G'(\eta ')=G(\eta)\;.
\end{equation}
In particular, the infinitesimal form of the generalized canonical
transformations is
\begin{equation}\label{a20}
\eta'=\eta+\delta\eta\;,\;\;\; \delta\eta=\left\{\eta,\delta W\right\}
^{(\eta,\Lambda)}\;.
\end{equation}

Let us suppose now that the classical mechanics in question has
equations of motion of the form
\begin{equation}\label{a21}
\dot{\eta}=\left\{\eta,H\right\}^{(\eta,\Lambda)}\;,
\end{equation}
i.e. the hamiltonian equations of motion, but with a generalized
Poisson bracket. How they
are transformed under the generalized canonical transformations
(\ref{a20}) ? The result is
\begin{equation}\label{a22}
\dot{\eta}'=\left\{\eta',H'\right\}^{(\eta',\Lambda)}\;,
\;\;\;H'(\eta')=H(\eta)+\frac{\partial \delta W}{\partial t}\;.
\end{equation}
It means that the equations (\ref{a21}) are forminvariant under
the generalized canonical transformations, only Hamiltonian is
changed, similar to the usual case of the canonical
transformations and hamiltonian
equations of motion with the ordinary Poisson bracket. To see this,
one has to calculate the time derivative of $\eta'$,  using (\ref{a21}),
\[
\dot{\eta}'=\left\{\eta+\delta\eta,H\right\}^{(\eta,\Lambda)}+
\left\{\eta,\frac{\partial \delta W}{\partial t}\right\}^{(\eta,\Lambda)}=
\left\{\eta+\delta\eta,H+\frac{\partial \delta W}{\partial t}
\right\}^{(\eta,\Lambda)}\;.
\]
Taking into account (\ref{a21}),(\ref{a20}), and (\ref{a19}), we obtain just
equations (\ref{a22}).

If a physical quantity is represented by a function $F(\eta)$ in
the variables $\eta$ then in the primed variables (\ref{a10}) it will be
represented by a function $F'(\eta ')$, which is related to the former
one by the eq. $F'(\eta ')=F(\eta)$.  In
the infinitesimal form it results in $F'(\eta)=F_{\delta W}(\eta)$,
according the eq.(\ref{a22}),
\begin{equation}\label{a23}
F_{\delta W}(\eta)=F(\eta)+\delta F(\eta)\;,\;\;\;
\delta F(\eta)=\left\{\delta W,F
\right\}^{(\eta,\Lambda)}\;.
\end{equation}

Variations of the phase variables in course of the time evolution
(\ref{a20}) can also be considered as a generalized canonical
transformation. Namely, let $\eta$ are the phase variables at a time
instant $t$ and $\eta_0$ are ones at the time instant $t=0$.
Then $\eta$ are some functions of $\eta_0$ and of $t$ as a
parameter, $\eta=\varphi(\eta_0,t)$. One can see that the
transformation from $\eta_0$ to $\eta$ is a generalized canonical
transformation. Moreover, this transformation can be formally written
explicitly. Indeed, considering for simplicity time independent
Hamiltonians only, one can see that the solution of the Cauchy problem
for the equation (\ref{a20}), with the initial data $\eta_0$ at $t=0$,
has the form
\begin{equation}\label{a24}
\eta=e^{\check{H}t}\eta_0\;,
\end{equation}
where the operator $\check{H}$ is defined by its action on functions
$F(\eta_0)$ of
$\eta_0$ as $\check{H}F(\eta_0)=\left\{F(\eta_0),H(\eta_0)
\right\}^{(\eta_0,\Lambda)}$. Because of the transformation
(\ref{a24}) is the generalized canonical transformation (see
(\ref{a10})) with the
generating function $H(\eta_0)$, one has only prove
that (\ref{a24}) obeys the equation of motion (\ref{a20}). Taking the
time derivative from (\ref{a24}), one gets
\begin{equation}\label{a25}
\dot{\eta}=\check{H}e^{\check{H}t}\eta_0=\left\{e^{\check{H}t}\eta_0,
H(\eta_0)\right\}^{(\eta_0, \Lambda)}\;.
\end{equation}
Using (\ref{a12}), one can verify that
\begin{equation}\label{a26}
H(e^{\check{H}t}\eta_0)=e^{\check{H}t}H(\eta_0)=H(\eta_0)\;.
\end{equation}
Substituting (\ref{a26}) into (\ref{a25}) and taking into account the
property (\ref{a19}), one obtains
\[
\dot{\eta}=\left\{e^{\check{H}t}\eta_0,
H(e^{\check{H}t}\eta_0)\right\}^{(\eta_0, \Lambda)}=\left\{\eta,
H(\eta)\right\}^{(\eta, \Lambda)}\;,
\]
what proves our affirmation.

\section{D-transformations}

Now we are going to apply the previous consideration to theories
with constraints, namely, with second-class constraints.

Let us consider a theory with second-class constraints
$\Phi=(\Phi_{l}(\eta))$, in hamiltonian formulation, described by
phase variables $\eta^A,\;\;A=1,\ldots,2n$, half of which are
coordinates $q$ and half
are moments $p$, so that $\eta^A=(q^a,p_a),\;\; A=(\zeta,a),\;\;
\zeta=1,2,\;\; a=1,\ldots,n$. An important object in such theories is
the Dirac bracket between two functions $F(\eta)$ and $G(\eta)$,
\begin{equation}\label{b1}
\{F,G\}_{D(\Phi)}=\{F,G\}-\{F,\Phi_l\}\{\Phi
,\Phi\}^{-1}_{ll'}\{\Phi_{l'},G\}\;.
\end{equation}
It is easy to see that the Dirac bracket is a particular case of the
generalized Poisson bracket (\ref{a1}),
\begin{equation}\label{b2}
\{F,G\}_{D(\Phi)}=\{F,G\}^{(\eta,\Lambda)}\;,
\end{equation}
with
\begin{equation}\label{b3}
\Lambda^{AB}=E^{AB}-\{\eta^A,\Phi_l\}\{\Phi
,\Phi\}^{-1}_{ll'}\{\Phi_{l'},\eta^B\}=\{\eta^A,\eta^B\}_{D(\Phi)}\;.
\end{equation}
If so, then one can consider the generalized canonical transformations for
such a generalized Poisson bracket. This
special but important case of the generalized canonical
transformations we will call D-transformations. Thus, by the
definition, D-transformations $\eta\rightarrow\eta'$ preserve the
forminvariance of the
Dirac bracket\footnote{A prime above the Dirac bracket
in (\ref{b4}) means that the latter is calculated in the primed variables.},
\begin{equation}\label{b4}
\{F,G\}_{D(\Phi)}=\{F',G'\}'_{D(\Phi)}\;.
\end{equation}
As we will see further, in theories with second-class constraints,
D-transformations play the same role which play   canonical transformations in
theories without constraints.

An explicit form of D-transformations connected continuously with the
identical transformation can be extracted from (\ref{a10}) and (\ref{b2}),
\begin{equation}\label{b5}
\eta '=e^{\check{W}}\eta\;,\;\;\;\check{W}F(\eta)=\left\{F ,W
\right\}_{D(\Phi)}\;,
\end{equation}
and in the infinitesimal form
\begin{equation}\label{b6}
\eta'=\eta+\delta\eta\;,\;\;\; \delta\eta=\left\{\eta,\delta W\right\}
_{D(\Phi)}\;,
\end{equation}
where $W(\eta)$ is a generating function of the D-transformation.

One can see that D-transformations differ from canonical ones only by
terms proportional to constraints. Indeed, the variation $\delta \eta$
under the D-transformation can be written as
\begin{equation}\label{b7}
\delta\eta=\left\{\eta,\delta W\right\}_{D(\Phi)}=\left\{\eta,\delta
W'\right\}+\{\Phi\}\;,
\end{equation}
where
\[
\delta W'=\delta W-\Phi_l\{\Phi,\Phi\}^{-1}_{ll'}\{\Phi_{l'},\delta
W\}\;,
\]
and $\{\Phi\}$ accumulates terms proportional to constraints, or terms
which vanish on the constraint surface.

As is known \cite{Dir2} equations of motion for a theory with second-class
constraints can be written in the form
\begin{eqnarray}
&&\dot{\eta}=\{\eta,H\}_{D(\Phi)}\;,\label{b8}\\
&&\Phi(\eta)=0\;.\label{b9}
\end{eqnarray}
They consist of two groups of equations, hamiltonian equations
(\ref{b8}) with the Dirac bracket, which is in the same time
a generalized Poisson bracket, and equations of constraints (\ref{b9}).
Using the previous section consideration, one can say that the
equations (\ref{b8}) are forminvariant under the D-transformations.
It turns also out that the equations of constraints (\ref{b9}) are
forminvariant under the D-transformations. Indeed, let $\Phi'(\eta')=0$
are equations of constraints in  variables $\eta'$, connected with
$\eta$ by a D-transformation, then the
relations
\begin{equation}\label{b10}
\Phi'(\eta')=\Phi(\eta)
\end{equation}
have to hold. One can consider these relations as  functional equations for the
functions $\Phi'$. It is easily to verify that they have a solution
$\Phi'=\Phi$. Indeed, consider the functions $\Phi(\eta')$. Using the
formula (\ref{a12}) and a well known property of the Dirac bracket:
$\left\{F ,\Phi_{l}\right\}_{D(\Phi)}=0$ for any function $F(\eta)$
and any constraint $\Phi_{l}$, we get
\begin{equation}\label{b11}
\Phi(\eta')=e^{\check{W}}\Phi(\eta)=\Phi(\eta)\;.
\end{equation}
That means that the constraints surface $\Phi(\eta)=0$ after the
D-transformation can be described  by the same functions, i.e. by
the equations $\Phi(\eta')=0$.

Thus, equations of motion of theories with second-class constraints
are forminvariant under the D-transformations. Namely, the equations
(\ref{b8}) and (\ref{b9}) have the following form after the
D-transformation (\ref{b6}):
\begin{equation}\label{b12}
\dot{\eta}'=\{\eta',H'\}'_{D(\Phi)}\;,\;\;\;
\Phi(\eta')=0\;,\;\;\; H'(\eta')=H(\eta)+\frac{\partial \delta W}
{\partial t}\;,
\end{equation}
or
\begin{equation}\label{b13}
\dot{\eta}=\left\{\eta,H_{\delta W}+\frac{\partial \delta W}{\partial t}
\right\}_{D(\Phi)}\;,\;\;\; \Phi(\eta)=0\;,
\end{equation}
and the physical quantities $F$ are described by the functions
$F_{\delta W}(\eta)$, see (\ref{a23}),
\begin{equation}\label{b14}
F'(\eta)=F_{\delta W}(\eta)=F(\eta)+\{\delta W,F\}_{D(\Phi)}\;,
\end{equation}

In the special canonical
variables $(\omega,\Omega)$, in which equations of constraints have
a simple form $\Omega=0$ (see \cite{GiT1,GiT2}), and the Dirac bracket reduces
to the Poisson one in the variables $\omega$, so that the latter are
physical variables on the constraints surface, D-transformations have
a simple meaning: they are canonical transformations in the sector of
physical variables $\omega$ with no change of variables $\Omega$. It is natural
because the D-transformations do not change the form of constraints.

\section{Quantum implementation of D-transformations}

One can ask a question: which kind of transformations in
quantum theory corresponds to  D-transformations in classical theory?
It is easy to see that these are unitary transformations and vice
versa: unitary transformations in a
quantum theory with constraints induce in a sense D-transformations
in the corresponding classical theory. From this point of view
D-transformations in theories with constraints
play also the role similar to one of the canonical transformations in
theories without
constraints. To prove this affirmation we have to remember that
in a classical theory D-transformations are transformations of
trajectories-states of the theory. Thus, if to speak literally, some
transformations of quantum states-vectors in a Hilbert
space, have to correspond them in a quantum
theory.

Let us have a classical theory with second-class constraints, which is
described by the equations of motion (\ref{b8},\ref{b9}). Its canonical
quantization \cite{Dir2,GiT1} consists formally in a transition from
the classical variables $\eta$ to quantum operators
$\hat{\eta},\;P(\hat{\eta}^A)=P(\eta^A)
=P_A$, which obey the
operator relations\footnote{Via $[\hat{A},\hat{B}\}$ we denote a
generalized commutator of two
operators $\hat{A}$ and $\hat{B}$, with definite parities
$P(\hat{A})$ and $P(\hat{B}),\;\;[\hat{A},\hat{B}\}=\hat{A}\hat{B}-
(-1)^{P(\hat{A})P(\hat{B})}\hat{B}\hat{A} $. An overline
with a hat, above a classical function $A(\eta)$, here and further
means a certain rule of
correspondence between the function and the corresponding quantum
operator $\hat{A},\;\;
\hat{A}=\hat{\overline{A(\eta)}}$. The former is in this case the
symbol of the operator \cite{Ber}.
 A choice of this rule is not important in our considerations.}
\begin{equation}\label{c1}
[\hat{\eta}^A,\hat{\eta}^B\}=i\hbar\hat{\overline{\{\eta^A,\eta^B\}}}_{D(\Phi)}
=i\hbar
\hat{\overline{\Lambda^{AB}(\eta)}},\;\;\;\hat{\overline{\Phi (\eta)}}=0,
\end{equation}
and which suppose to be realized in a Hilbert space ${\cal R}$ of vectors
$|\Psi>$. Then one has to assign operators $\hat{F}$ to all the
physical quantities $F$, which are described in the
classical theory by the functions $F(\eta)$, using a certain
correspondence rule,  $\hat{F}=\hat{\overline{F(\eta)}}$. The time
evolution of the state vectors is
defined by the quantum Hamiltonian $\hat{H}=\hat{\overline{H(\eta)}}$,
according the Schr\"odinger equation
\begin{equation}\label{c2}
i\hbar\frac{\partial |\Psi>}{\partial t}=\hat{H}|\Psi>\;.
\end{equation}

Let us consider a unitary transformation of the state vectors,
$|\Psi>\rightarrow |\Psi '>=\hat{U}|\Psi>$, where $\hat{U}$ is some
unitary operator, $\hat{U}^+\hat{U}=1$, which one can write in the
form
\begin{equation}\label{c3}
\hat{U}=\exp\left\{-\frac{i}{\hbar}\hat{W}\right\} \; ,
\end{equation}
where $\hat{W}$ is a hermitian operator, $\hat{W}^+=\hat{W}$, further
called quantum generator of the transformation. In the
infinitesimal form ($\hat{W}\rightarrow \delta \hat{W}$), simplifying
the consideration, $|\Psi'>=|\Psi>+\delta |\Psi>,\;\; \delta |\Psi>=
-\frac{i}{\hbar}\delta \hat{W} |\Psi>$.

One can find a variation of operators of physical quantities from the
condition \break $<\Psi|\hat{F}|\Psi>=<\Psi '|\hat{F}'|\Psi '>$, which
results in
\begin{equation}\label{c4}
\hat{F}'=\hat{F}_{\delta W}=\hat{U}\hat{F}\hat{U}^+ =\hat{F}+\delta
\hat{F},\;\;\;\delta \hat{F}=-\frac{i}{\hbar}[\delta \hat{W},\hat{F}]\;.
\end{equation}
If $\delta W(\eta)$ is a symbol of the operator  $\delta \hat{W},\;\;
\delta \hat{W}=\hat{\overline{\delta W}}\eta)$ and
$F(\eta)$ is one of the operator $\hat{F}$ (the classical function
which describes the physical quantity in the variables $\eta$),
$\hat{F}=\hat{\overline{F(\eta)}}$, then
it follows from the eq. (\ref{c1})
\begin{equation}\label{c5}
\delta \hat{F}=\hat{\overline{\{\delta W,F\}}}_{D(\Phi)}+o(\hbar)\;.
\end{equation}
Remembering the formula (\ref{b14}), one can write
\begin{equation}\label{c6}
\hat{F}_{\delta W}=\hat{\overline{F_{\delta W}(\eta)}}+o(\hbar)\;.
\end{equation}
Thus, operators of physical quantities,  transformed in
course of a unitary
transformation, have as their symbols initial classical functions
transformed by a D-transformation, with
the generating function  being a classical  symbol of the quantum generator of
the unitary transformation.

The Schr\"odinger equation for transformed vectors can be derived from
the eq. (\ref{c2}) and has the form
\begin{equation}\label{c7}
i\hbar\frac{\partial |\Psi '>}{\partial t}=\hat{H} '|\Psi '>\;,\;\;\;
\hat{H}'=\hat{H}_{\delta W}+\frac{\partial}{\partial
t}\hat{\overline{\delta W}}\;.
\end{equation}
Thus, the time evolution of the state vectors after the unitary
transformation is governed by a quantum Hamiltonian with the classical
symbol
\begin{equation}\label{c8}
H'(\eta)=H_{\delta W}(\eta)+\frac{\partial \delta W(\eta)}{\partial t}+
o(\hbar)\;.
\end{equation}
That fact and the eq.(\ref{c1}) allow one  to see that the classical
limit of the quantum theory after the unitary transformation
(\ref{c3}) is described by the equations
(\ref{b13}) and therefore corresponds to the D-transformed classical
theory with
the generating function, which is a classical symbol of the quantum
generator of the unitary transformation. In the same way one can
prove an inverse statement: if we have a classical theory and its
D-transformed formulation, then quantum versions of both theories are
connected by an unitary transformation. Besides, the classical
generating function of the D-transformation and
the quantum generator of the unitary transformation are connected in
the above mentioned menner.

Consider now the generating functional $Z(J)$ of Green's functions  for
theory with second-class constraints in the form of hamiltonian path integral
and  a behavior of the latter under the D-transformations. Such an
integral can be written in the form
\begin{equation}\label{c9}
Z(J)=\int\exp\left\{\frac{i}{\hbar}S_J(\eta)\right\}{\bf {\cal D}}\eta\;,
\end{equation}
where
\[
S_J(\eta)=\int[p_a\dot{q}^a-H_J(\eta)]dt,\;\; H_J(\eta)=H(\eta)
+J_A\eta^A\;,
\]
is the classical action with sources, $J_A(t)$ are
sources to the variables $\eta^A(t), \;\;P(J_A)=P(\eta^A)=P_A$, and
the measure ${\cal D}\eta$ has the form \cite{Fad,Fra},
\begin{equation}\label{c10}
{\bf {\cal D}}\eta={\rm Sdet}^{1/2}\{\Phi,\Phi\}\delta(\Phi)D\eta \;,
\end{equation}
with ${\rm Sdet}\{\Phi,\Phi\}$ denoting the superdeterminant of the
matrix $\{\Phi_l,\Phi_m\}$.

As is known, if a change of variables $\eta'=\eta'(\eta)$ is a canonical
transformation, then $|{\rm Ber}\;\eta'(\eta)|=1$, where ${\rm
Ber}\eta'(\eta)$ is Berezinian \cite{Ber} of the change of variables,
${\rm Ber}\;\eta'(\eta)={\rm S det}\partial_r \eta'^A/\partial \eta^B$. In
particular, for infinitesimal canonical transformations
$\eta '=\eta+\delta \eta,\;\; \delta\eta=\{\eta,\delta W\},\;\; {\rm
Ber}\;\eta'(\eta)=1$. In case of theories without constraints, the
measure ${\bf {\cal D}}\eta$  (\ref{c10}) reduces to $D\eta$ and is
invariant under canonical transformations, but in theories with
constraints it is not. However, this measure is invariant under
D-transformations,
\[
{\cal D}\eta'={\cal D}\eta\;,
\]
which confirms ones again that the latter play the
role of canonical transformations in theories with constraints. To see
this one can use a relation \cite{GiT2},
\begin{equation}\label{c11}
\left.{\rm
Sdet}^{1/2}\{\Phi,\Phi\}\delta(\Phi)\right|_{\eta\rightarrow
\eta'(\eta)}
{\rm Ber}\;\eta'(\eta)={\rm Sdet}^{1/2}\{\Phi,\Phi\}\delta(\Phi)\;,
\end{equation}
where $\eta'=\eta+\{\eta,\delta W\}_{D(\Phi)}$.

The invariance  of the measure (\ref{c10}) under
D-transformations, induces an invariance of the integral (\ref{c9})
under the transformations of the action $S_J(\eta)$,
\begin{equation}\label{c12}
S_J(\eta)\rightarrow S'_J(\eta)=S_J(\eta'(\eta))=S_J(\eta)+\delta
S_J(\eta)\;,
\end{equation}
where $\eta'(\eta)$ are D-transformations,
\[
Z(J)=\int\exp\left\{\frac{i}{\hbar}S_J(\eta)\right\}{\bf {\cal D}}\eta=
\int\exp\{\frac{i}{\hbar}S'_J(\eta)\}{\bf {\cal D}}\eta\;,
\]
or
\begin{equation}\label{c13}
\int\delta S_J(\eta) \exp\left\{\frac{i}{\hbar}S_J(\eta)\right\}
{\bf {\cal D}}\eta=0\;.
\end{equation}
It is enough to know $\delta S_J(\eta)$ on the
constraints surface, because of the integration in (\ref{c13}) is only going
over this surface due to the $\delta$-function in the measure
(\ref{c10}). Taking into account the
representation (\ref{b7}), one can find an expression for
$\delta S_J(\eta)$ on the constraints surface,
\begin{equation}\label{c14}
\left.\delta S_J(\eta)\right|_{\Phi=0}=\left.(p\delta q-\delta W)
\right|_{t_{in}}^{t_{out}}+\int\left[\frac{\partial }{\partial t}\delta
W-\{H_J,\delta W\}_{D(\Phi)}\right]dt\;.
\end{equation}
In field theory usually
$t_{in,out}\rightarrow \pm \infty$ and trajectories of integration
vanish at these time limits. Considering D-transformations, which do
not change this property, one gets
\begin{equation}\label{c15}
\int \left[\int\left(\frac{\partial }{\partial t}\delta
W-\{H_J,\delta W\}_{D(\Phi)}\right)dt\right]\exp \left\{\frac{i}{\hbar}
S_J(\eta)\right\}{\bf {\cal D}}\eta=0\;.
\end{equation}
This relation can be used to obtain different kinds of equations for
generating functional and therefore for Green's functions. For
example, let us consider
D-transformations with  two types of generating functions: $\delta W
=\epsilon_A\eta^A$, and
$\delta W=\zeta_l\Phi_l(\eta)$, with arbitrary ``small'' time
dependent functions $\epsilon_A(t)$ and $\zeta_l(t)$. Using these $
\delta W$ in eq.
(\ref{c15}), we get two relations
\begin{eqnarray}\label{c16}
&&\int\left[\dot{\eta}^A-\{\eta^A,H_J\}_{D(\Phi)}\right]\exp
\left\{\frac{i}{\hbar}
S_J(\eta)\right\}{\bf {\cal D}}\eta=0\;,\nonumber \\
&&\int \Phi_l(\eta)\exp \left\{\frac{i}{\hbar}
S_J(\eta)\right\}{\bf {\cal D}}\eta=0\;,
\end{eqnarray}
which can be rewritten in the form of Schwinger equations for the
functional $Z(J)$,
\begin{equation}\label{c17}
\left[\dot{\eta}^A-\{\eta^A,H_J\}_{D(\Phi)}\right]_{\eta
\rightarrow \frac{\delta_l}{\delta (iJ)}}Z(J)=0\;,\;\;\;
\Phi\left(\frac{\delta_l}{\delta (iJ)}\right)Z(J)=0\;.
\end{equation}

\section{Remarks}
Thus, we demonstrated that in theories with second-class constraints
D-transformations play the usual role of canonical ones. In fact, in
our books \cite{GiT1} we have  already used infinitesimal
D-transformations for technical reasons, but at that time we did not fully
realize  their special role.

Author thanks Prof. Igor Tyutin for  discussions and helpful remarks
and Prof. Jose Frenkel for discussions and friendly support.


\begin{thebibliography}{99}
\bibitem{LaL}L.D.Landau and E.M.Lifshitz, {\em Mechanics} (Nauka,
Moscow, 1973)
\bibitem{Dir1}P.A.M.Dirac, {\em The Principles of Quantum Mechanics}
(4 th. ed., Oxford Univ. Press, Oxford, 1958)
\bibitem{Wey}H.Weyl, {\em The Theory of Groups and Quantum
Mechanics}  (2nd. ed., Dover, New York, 1950, 1931)
\bibitem{Wit}B.DeWitt, Phys.Rev. {\bf 85} (1951) 653
\bibitem{Itz}C.Itzykson, Commun. Math. Phys. {\bf 4} (1967) 92
\bibitem{MoQ}M.Moshinsky and C.Quesne, J. Math. Phys. {\bf 12}
(1971) 1772
\bibitem{And}A.Anderson, Annals of Phys. {\bf 232} (1994) 292
\bibitem{Dir2}P.A.M.Dirac, {\em Lectures on Quantum Mechanics} (New York:
Yeshiva University, 1964)
\bibitem{GiT1}D.M.Gitman and I.V.Tyutin, {\em Canonical Quantization of Fields
with Constraints} (Nauka, Moscow, 1986);
{\em Quantization of Fields  with Constraints}  (Springer-Verlag, 1990)
\bibitem{GiT2}D.M.Gitman and I.V.Tyutin, Soviet Phys. Journ. {\bf 26} (1983)
423; {\em Structure of Gauge Theories in Lagrangian and Hamiltonian
Formalisms} in {\em Quantum Field Theory and Quantum Statistics} Vol.1, p.143,
ed. by I.A.Batalin, C.J.Isham, G.A.Vilkovisky (Adam Hilger, Bristol, 1987)
\bibitem{Ber}F.A.Berezin, {\em The Method of Second Quantization}
(Nauka, Moscow, 1965); {\em Introduction to Algebra and Analysis with
Anticommuting Variables} (Moscow State University press, Moscow,
1983); {\em Introduction to Superanalysis} (D.Reidel, Dordrecht, 1987)
\bibitem{Fad}L.Faddeev, Theor. Mat. Fiz. {\bf 1} (1969) 3
\bibitem{Fra}E.S.Fradkin,
Proc. of Tenth Winter School of Theor. Phys. in Karpacz, Acta Univ.
Wratisl. {\bf N207} (1973)

\end{thebibliography}
\end{document}